\def\year{2022}\relax
\documentclass[letterpaper]{article} 
\usepackage{aaai22}  
\usepackage{times}  
\usepackage{helvet}  
\usepackage{courier}  
\usepackage[hyphens]{url}  
\usepackage{graphicx} 
\usepackage{natbib}  
\usepackage{caption} 
\DeclareCaptionStyle{ruled}{labelfont=normalfont,labelsep=colon,strut=off} 
\usepackage{subfig}
\usepackage{algorithm}
\usepackage{algorithmic}

\usepackage{xcolor}

\usepackage{amsmath,amssymb,bm}
\usepackage{color, colortbl}
\definecolor{Gray}{gray}{0.9}
\usepackage{multicol}
\usepackage{lscape}

\usepackage{newfloat}
\usepackage{listings}
\floatstyle{ruled}
\newfloat{listing}{tb}{lst}{}
\floatname{listing}{Listing}
\title{BotBuster: Multi-platform Bot Detection Using A Mixture of Experts}
\author{
    Lynnette Hui Xian Ng\footnote{Corresponding Author}, Kathleen M. Carley\textsuperscript{\rm 1}
}
\affiliations{
    \textsuperscript{\rm 1}CASOS, Institute for Software Research, \\ Carnegie Mellon University\\
    \{huixiann,carley\}@andrew.cmu.edu
}

\begin{document}

\maketitle

\begin{abstract}
Despite rapid development, current bot detection models still face challenges in dealing with incomplete data and cross-platform applications. 
In this paper, we propose BotBuster, a social bot detector built with the concept of a mixture of experts approach. Each expert is trained to analyze a portion of account information, e.g. username, and are combined to estimate the probability that the account is a bot. 
Experiments on 10 Twitter datasets show that BotBuster outperforms popular bot-detection baselines ($\bar{F1}=73.54$ vs $\bar{F1}=45.12$). 
This is accompanied with F1=60.04 on a Reddit dataset and F1=60.92 on an external evaluation set.
Further analysis shows that only 36 posts is required for a stable bot classification. Investigation shows that bot post features have changed across the years and can be difficult to differentiate from human features, making bot detection a difficult and ongoing problem.
\end{abstract}

\section{Introduction}
Social bots accounts are partially or fully controlled by algorithms. Social bot detection have been extensively studied for more than a decade and many machine learning frameworks have sprung up to address this problem \cite{10.1145/3409116}.

Despite this progress, two important issues remain: dealing with incomplete data and a multi-platform bot detector.
Bot detectors rely on account features from content information to user metadata to perform a prediction.
However, during fast-moving events like elections or protests, data collection on accounts is often incomplete. It is near impossible to perform an extensive collection of all fields required by most bot detection algorithm, especially when there are at least a million accounts. 
Other times, it is simply impossible to collect all required data: the account may have been suspended or turned protected, forcing analysts to rely on available data or previously collected historical data.
A popular method to fill incomplete data is to make missing values zero, but that may affect the prediction as zero or null are valid values that do occur in the data. 
Additionally, most bot detectors are currently tuned for the Twitter platform, leaving other social media platforms vulnerable.

In this work, we advance the problem of bot detection modules through a mixture of experts architecture that can handle incomplete account information. 
Given an account's information of {username, screenname, description, user metadata, posts}, where some information in this set may be missing, our goal is to classify the account into one of two classes: {bot, human}. Each expert is trained to analyze one pillar of data and are combined to estimate the probability that the account is a bot.
The proposed solution, \textbf{BotBuster}, aims to overcome the limitations of requiring the entire set of account information for effective bot prediction and expand bot prediction to multi-platforms.

\paragraph{Contributions} Our contributions include:
\begin{enumerate}
    \item We introduce the concept of a mixture of experts approach to bot prediction, which overcomes the limitation of requiring data for all features and make a decent prediction given available data. The proposed BotBuster reaches state-of-the-art performance ($\bar{F1}$=72.23), outperforming baselines ($\bar{F1}$=49.61, 40.63).
    \item We extend bot prediction from one platform to multiple platforms, incorporating bot prediction across Twitter and Reddit. BotBuster stabilizes after 36 posts and performs robustly on an external validation set (F1=60.92).
    \item Finally, we investigate the difficulty of the bot detection task through the blurred distribution of features between bots/humans, and the changing distribution of features across the years.
\end{enumerate}

\section{Related Work}

\paragraph{Bot Detection}
Common supervised Twitter bot-detection mechanisms are feature-based, using account details such as tweet frequency \cite{yang2020scalable}, tweet content \cite{beskow2018bot}, temporal features \cite{chavoshi2016debot} and network features \cite{yang2019arming}. 
Popular bot-detection libraries are: Botometer \cite{varol2017online}, BotHunter \cite{beskow2018bot}. 
Botometer uses a supervised ensemble classification based on user, tweet and network features extracted for each account. 
BotHunter uses a multi-tiered random forest method, with each tier making use of more features as before, from content to user to network features.

However, Botometer queries Twitter live, meaning it is unable to work with archived data, BotHunter requires the full feature set for prediction. We bridge this gap by building a bot prediction algorithm that can work with incomplete and historical data.


Bot detection has been widely studied on Twitter; it is less so on Reddit. Past work includes exploiting temporal and network information to detect political bots \cite{10.1145/2783258.2783294, 10.1145/3313294.3313386}.
While there have been research on cross-platform analysis of social bot trends \cite{mittos2020analyzing}, to the best of our knowledge, there has not been a bot prediction model that successfully combines bot prediction for Twitter and Reddit platforms.

\paragraph{Mixture of Experts}
Mixture of experts models have been used in ensemble learning: for translation \cite{peng2020mixture}, sequence learning \cite{shazeer2017outrageously} and text generation \cite{shen2019mixture}.
A close cousin is multi-task learning, where there are shared parameters among the tasks \cite{ruder2017overview}. This approach has been used in stance detection \cite{li-caragea-2021-multi} and offensive language detection \cite{benton2017multitask}.
To the best of our knowledge, no bot detection approach has harnessed this model for bot detection. We adapt these ideas from the natural language community for bot detection. 

\section{BotBuster Architecture}
\begin{figure*}
\centering
\includegraphics[width=1.0\textwidth]{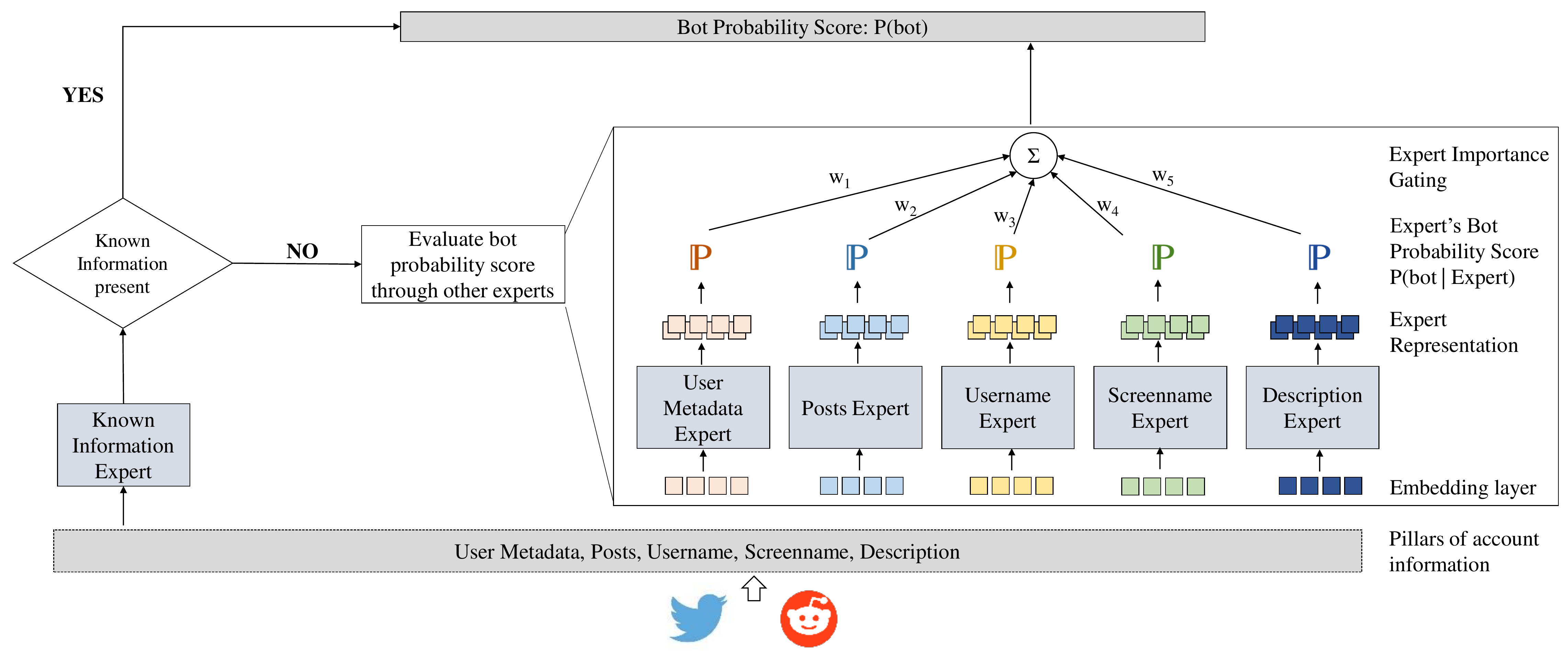} 
\caption{Diagram of the BotBuster Architecture.}
\label{fig:architecture}
\end{figure*}

Figure \ref{fig:architecture} illustrates our proposed BotBuster architecture flow.
Given the set of user account information: (user name, screen name, description, user metadata, posts), we first evaluate the bot probability score ($\bm{P}(\text{bot})$) through a Known Information Expert. If known information is present in the information pillars, $\bm{P}(\text{bot})$ is returned.

If known information is not present, $\bm{P}(\text{bot})$ is evaluated as a weighted sum of bot probabilities evaluated through 5 other experts, one for each information pillar.
Each expert takes in an expert-specific input representation, an
embedding vector of their corresponding information pillar. It returns a 64-d Expert Representation Vector and the Expert's Bot Probability Score ($P(bot|Expert)$), the probability of the account being a bot given the expert's information. The Expert Representation Vectors serve as inputs to a gating 
network which calculates a bot probability score, $\bm{P}(\text{bot})\in[0,1]$.
The experts are grouped by fields that are commonly retrieved together; should one of the fields the expert requires be absent, the expert is not activated. The gating weights change accordingly to the number of pillars present, accounting for incomplete data.

\paragraph{Known Information Expert} is the first gate which evaluates bot or not based on definite known information. If the signals from these definite known information are not present, the collected user data processes through the rest of the BotBuster architecture. 
The expert are activated under the following two conditions: (1)$\bm{P}(\text{bot})$=0 if the ``is\_verified" field is True for Twitter accounts, and (2)$\bm{P}(\text{bot})$=1 if the accounts has the word ``bot" in the extracted user features  (eg. ``xxUpdateBot").
Although the final probability of the account has been determined through these definite signals, these information are still used to train the other experts.

\paragraph{Username, Screenname, Description Experts} are modelled similarly and trained on their respective data pillars. Their input representation is a 768d-vector of BERT embeddings of their data pillars, which is then fed into a pre-trained BERT model with a Multi-Layer Perceptron (MLP) network for fine-tuning that outputs a 64d Expert Representation for the gating network and $P(bot|Expert)$.

\paragraph{User Metadata Expert} takes in user metadata as a vector into a 4-layer MLP with a dense layer. It returns a 64d Expert Representation for the gating network and $P(bot|Expert)$.

\paragraph{Post Expert} derives $P(bot|Expert)$ and a 64d-representation from post texts. We test two types of post experts: account-level and post-level. 

In the pre-processing step, we first filter for posts that are origin posts, ie. posts that are not quotes or retweets. This step gives us an insight into the post information and linguistic style originating from the agent. We then remove any hashtags, @-mentions, URLs and stop words. We derive linguistic values from the post text as features which are described in \textit{Feature Engineering}.

A post consists of two portions: post texts and post metadata (eg., retweet count, like count etc. and derived linguistic values).
For post texts, we tokenize sentences using the default tokenizer of the BERT language model to obtain sequence embeddings. 
For post metadata and derived linguistic values, we construct a normalized float vector each. 

\textbf{Account-Level Post Expert.}
In this variation, we construct two sub-post experts: one representing the post texts and another representing post metadata. We feed the post texts expert into a BERT model and fine-tuned it with a dense layer. We feed the post metadata expert into a 4-layer MLP followed by a dense layer. We then combine the output of the two sub-posts experts in a uniform manner to obtain the probability $P(bot|expert)$. We concatenate the 32d-representation of both sub-post expert to form a 64d- post Expert Representation for use in the gating network.

\textbf{Post-Level Post Expert.} The post-level post expert is a joint model in a Siamese network structure.
We first feed the post text into a BERT model and fine-tuned it with a dense layer to obtain an intermediate embedding of the post text. 
We next feed post metadata into a 4-layer MLP and obtain an intermediate embedding of post metadata.
We concatenate the outputs of the intermediate embeddings, then trained a dense layer on top, to output the required information.

For each type of post expert, we experimented with and without the inclusion of derived post linguistic values as features. 
The account-level experts aggregates information from multiple posts as input, while the post-level expert uses information from each post singularly. As such, we do not construct a mixture that combines both types of post experts which will use information from a post twice.



\paragraph{Expert Importance Gating}
The ideal BotBuster input is all five pillars of data: (User Metadata, Posts, Username, Screenname, Description). Other combinations of data input include: (Username, Posts), (Screenname, Posts, Username) etc., in which one pillar of data is missing. In total, there are $5!=120$ combinations, each of which has a different weight distribution to the experts.

We adapt the approach from  \citet{peng2020mixture} to assign weights to each expert for a combined ensemble model. We first concatenate the 64-dimension Expert Representation from each expert and average them in the sequence direction, producing one representation per expert. Then we form an input vector that concatenates these average representations as input into the gating network. The gating network is a two-layer MLP, interspersed with \textit{tanh} activation layers. It has a final \textit{softmax} layer that normalizes the layer weights to output an $n$-dimensional vector that sums to 1, where $n$ is the number of input data pillars. The magnitude of each vector element represents the relative strength of the experts.

We trained this gating network to produce weights for all combinations of data input. For each combination, we run this training for three times and take the average weights. 
This Expert Importance Gating can be seen as a learned probability distribution over the experts, assigning more responsibility to the experts that are more important in bot detection to contribute more to $\bm{P}(\text{bot})$.

\section{Datasets}
We trained and tested our models on 10 public Twitter datasets from 2017 to 2020. These datasets are human annotated with bot/human labels and are publicly available on a bot repository  \footnote{\url{http://botometer.org/botrepository}}. 
The datasets range from accounts collected from the US elections to bots that were manually discovered on Twitter, covering a wide range of bot activities. 
We hydrate these accounts with the Twitter V2 REST API in July 2021, collecting user information and the latest 40 tweets per user.
The bot repository provides a user data file of some user metadata  of the accounts at the time of dataset creation. For some accounts, we are unable to retrieve information as they have been suspended by Twitter or became a protected accounts. We use the bot repository's user data file to enrich account information with data gaps.

We also construct an additional dataset for Reddit.
Unlike Twitter accounts, user information for Reddit accounts only consists of user names, providing us without information for screen name, posts and user metadata. 
We select the 500 highest ranked ``bad bot" in BotRank\footnote{\url{http://botrank.pastimes.eu}}, a crowdsourced bot curation list that provides a good/bad bot ranking based on community approval \cite{trujillo-etal-2021-echo}.

For Reddit humans, we collected users from 5 subreddits that generally require conscious writing and thought and manually verified the users are likely to be humans.  
We then use the Pushshift Reddit API \cite{baumgartner2020pushshift} to collect user information and the latest 20 posts of each account's timeline.

Table \ref{tab:datasets} provides a dataset overview and indicates the fields with incomplete data.
An ontology map unifies the field names to account for different naming conventions across social media platforms.

\subsection{Feature Engineering}
\label{sec:featureeng}
We perform feature engineering to exploit account's information.
Table \ref{tab:features} lists the features.

\paragraph{Extracted User Features} describe the user account in terms of user biography information (username, screenname, description), user statistics (number of followers/following, total number of posts, listed count) and account indicators (verified, protected flag).

\paragraph{Extracted Post Features} are gathered from the platform's API, includes post texts and its statistics (number of retweets, likes, quotes, replies).

\paragraph{Derived Post Features}
BotBuster-2/4 also uses derived post features, which are linguistic characterization of the texts.
This builds on past studies that observed bots used simpler language than humans \cite{Uyheng2021} and differ in the use of pronouns.

\textbf{Sentiment} of the text is calculated using the Pysentimento library \cite{perez2021pysentimiento}. The library fine-tunes a BERT-based network on a Twitter dataset annotated with positive, negative and neutral sentiments.  

The \textbf{Flesch-Kincaid reading difficulty score} standardized by the U.S. Military gauges the ease of readability of a text by the number of words and syllabus of the words \footnote{reading difficulty = 0.39 $\times$ average words per sentence + 11.8 $\times$ average syllabus per word - 15.59}.

\textbf{EPA scores} are affective social identities represented via three values: Evaluation (good vs. bad), Potency (strong vs. weak) and Activity (active vs. passive). These scores are extracted from a dictionary developed using surveys in 2012-2014 \cite{smith2016mean}. 

\textbf{LIWC features} are derived from the Linguistic Inquiry and Word Count (LIWC) lexicon that summarizes emotional, cognitive, and structural components in a text \cite{doi:10.1177/0261927X09351676}. We use the 2015 version and focus on the time, affect, social, drives and pronouns components.

\section{Experiments}
\paragraph{Experimental Setup}
In order for the BotBuster to learn from the diverse dataset acquired, we used a merged training strategy by training a single model on all the training data.
Our model is implemented with Tensorflow. 
For training and hyperparameter tuning of each expert, we select accounts that have a value for that expert.
We partition the entire dataset into 80:10:10 train:validation:test ratio. 
There is a disproportionately larger number of bots in the dataset, so we use stratified sampling to ensure there is an equal proportion of bot/human accounts in each set.
After individual expert training, we perform joint training for the Expert Importance Gating network, in which we similarly partition the dataset into 80:10:10 with stratified sampling.

For all experiments, we use ADAM as the learning optimization with a learning rate of 0.001. We run for 20 epochs with a batch size of 32. For the username, screenname, description, user metadata and account-level post experts, we use the binary cross-entropy loss. For the post-level post expert, we use the binary cross-entropy loss from logits, in order to back propagate the gradients to the individual posts. 

\paragraph{Models} 
We perform experiments on four variations of the BotBuster architecture. In all variations, the structure of the user name, screen name, description and user metadata experts remain the same; we mainly vary the post expert between an account-level post expert and a post-level post expert and on the inclusion of derived post linguistic values. The four model variations are: 
\begin{enumerate}
    \item \textbf{BotBuster-1}: account information + account-level post experts
    \item \textbf{BotBuster-2}: account information + account-level post experts with derived post linguistic values
    \item \textbf{BotBuster-3}: account information + post-level post experts
    \item \textbf{BotBuster-4}: account information + post-level post experts with derived post linguistic values
\end{enumerate}

\section{Evaluation}
In the evaluation phase, we use the BotBuster model to perform predictions. With BotBuster, we classify accounts as a bot when $\bm{P}(\text{bot})\geq0.50$ and human otherwise, stemming from the use of a binary classification model. 

The micro-F1 score is adopted for model evaluation. It gives the same importance to each sample, accounting for the class imbalance situation. We ran each experiment three times and report the mean F1-score.


\paragraph{Individual Expert Baseline}
For each expert, we perform baseline comparisons of our neural network based formulation against common classifiers. For the text-based classifiers like user name, screen name, description and posts experts, we used a Term Frequency-Inverse Document Frequency (TF-IDF) vectorizer as a preprocessing step before fitting the data into the classifier. For the numeric data such as user metadata expert, we formulate the numeric data into a vector and normalize each numeric category before fitting the data into the classifier. For post experts that make use of the post text and derived values, we concatenate the TF-IDF vector with a vector of post derived values.

\paragraph{Ensemble Baseline}
For the mixture of experts formulation of BotBuster, we perform baseline comparisons to two commonly used commercial bot-detection algorithms: Botometer \cite{varol2017online} and BotHunter \cite{beskow2018bot}. Both algorithms provide a bot probability score between 0 and 1. These algorithms have been continually updated since their development in 2017-18. They have been widely used in social bot detection studies.

For Botometer, we queried their API\footnote{https://botometer.osome.iu.edu/} to retrieve the universal score returned as $\bm{P}(\text{bot})$. 
We set a threshold of 0.5 for bot classification (i.e., $\bm{P}(\text{bot})\geq0.5=\text{bot}$, as suggested by its authors \cite{varol2017online}.
For BotHunter, we obtained its library from its authors to perform bot detection.
We set a threshold of 0.7 for bot classification (i.e., $\bm{P}(\text{bot})\geq0.7=\text{bot}$ \cite{bothunterstability}.
We report the results of these baseline predictions for each dataset where it exists (i.e. the baseline algorithms return results). We also report the average of these baseline predictions across all datasets. 

\paragraph{Individual dataset analysis}
In this phase, we train and test the BotBuster model on each dataset one by one. Each dataset is partitioned into a 80:10:10 stratified split and trained singularly using the same architecture. We compare the predicted classification against the original bot/human annotations and report the results as \textbf{BotBuster-singular}.

\paragraph{Individual dataset analysis with combined data}
We also perform individual dataset analysis with a combined data setup for training. In the combined data setup, we aggregate the data instances from all datasets. 
During each dataset run, we partition 10\% of the targeted dataset for testing. We partition the remaining combined data into 90:10 train:validation, accounting for the distribution of bots/humans in the combined dataset. We use the remaining combined data for training. 
We compare the predicted classification against the original bot/human annotations and report the results as \textbf{BotBuster-Full}. 

We opted for this strategy as it mimics the baseline bot detection algorithms. These algorithms are continually updated by training on diverse datasets, including those selected in this work. As we cannot retrain these state-of-the-art models on separate train-test data subsets, although we would love to, we adopt a setup where we train on data from all datasets and test on data from each dataset separately.

\paragraph{Evaluation against baseline returns}
The two baseline algorithms do not return results for all queried accounts. 
We thus collect $\bm{P}(\text{bot})$ for accounts the algorithms returned without modifying any data fields.
Using the set of accounts in which the baselines returned $\bm{P}(\text{bot})$, we construct two sets of \textbf{BotBuster-Subset} based on the respective returns. We then compare our approach against the results returned by the baselines.
We also perform proportion z-tests to test whether the proportion of bots derived from BotBuster and baseline models are the same. We first identify bot accounts for each model according to their respective threshold values, and compare the proportions of bots identified by each model. We adopt this approach over significance testing via bot probability means as bot classifiers are typically used to identify bot/human instead of for the absolute score. 

\paragraph{Evaluation against external dataset}
We perform an external validation by predicting bot probability using the BotBuster model on an unknown dataset. 
We run the BotBuster-4 prediction model on the TwiBot-20 test data set, consisting of 1183 users \cite{feng2021twibot}. This dataset was constructed in 2020, whereas BotBuster was trained on datasets collected up to 2019.
This evaluation helps us ascertain the robustness of the architecture towards newer bot behavior and an unknown set of annotations.
For BotBuster, we used the tweets and user data provided in the dataset. 
We also ran the dataset through the baseline algorithms. For the Botometer baseline algorithm, we did not use the historical data provided; the API does an online pull of account data at the point of query.

\section{Results}
\paragraph{Individual expert training}
We present the performance of training each Expert on accounts with all pillars of account information in Table \ref{tab:indivexpertswclassical2}. The BotBuster formulation performs better than traditional classifiers, validating our approach. 
Individual experts perform as well as the ensemble results, showing that each expert can make a fairly accurate prediction of $P\text{(bot)}$. 


\begin{table*}[!htbp]
\centering
\begin{tabular}{|lllll|}
\hline
\textbf{Expert} & \textbf{Decision tree} & \textbf{Support Vector Classifier} & \textbf{Random Forest} & \textbf{BotBuster-formulation} \\ \hline 
User name & 54.96 & 54.96 & 54.96 & \textbf{62.01} \\ 
Screen name & 59.70 & 60.79 & 59.56 & \textbf{62.01} \\ 
Description & 63.19 & 60.79 & 60.79 & \textbf{69.33} \\ 
User metadata & 44.52 & 55.63 & 55.63 & \textbf{62.38} \\ 
Posts (account-level) & 61.48 & 59.34 & 59.34 & \textbf{64.22} \\ 
Posts (post-level) & 59.22 & 63.59 & 63.32 & \textbf{64.79} \\ 
Posts (post-level) + derived values & 67.44 & 71.25 & 72.15 & \textbf{74.50} \\ 
Posts (account-level) + derived values & 68.74 & 68.65 & 61.81 & \textbf{72.98} \\ 
\hline
\end{tabular}
\caption{Macro F1 accuracy scores of individual experts.}
\label{tab:indivexpertswclassical2}
\end{table*}

\paragraph{Individual dataset analysis}
In comparison to the BotHunter and Botometer baseline algorithms, BotBuster has better coverage: analyzing 100\% of Twitter accounts and also covers Reddit accounts (Table \ref{tab:percanalyzed}).
BotBuster analyzes $\bm{P}\text(bot)$ based on the available information. In contrast, BotHunter requires the presence of the entire feature set it is trained upon, while Botometer fetches information from Twitter in real-time, hence suspended accounts will return no results despite having previously data collected. 

\begin{table}[!htbp]
\centering
\begin{tabular}{|p{2.7cm}|p{1.5cm}|p{1.5cm}|p{1.4cm}|p{1.4cm}}
\hline
\textbf{Dataset} & \textbf{BotHunter} & \textbf{Botometer} & \textbf{BotBuster} \\ \hline
astroturf & 27.65 & 34.02 & 100 \\ 
botometer-feedback-2019 & 61.44 & 71.07 & 100 \\ 
botwiki-2019 & 90.34 & 92.90 & 100 \\ 
cresci-rtbust-2019 & 74.97 & 78.78 & 100 \\  
cresci-stock-2018 & 40.57 & 47.03 & 100 \\ 
gilani-2017 & 72.39 & 0 & 100 \\ 
midterm-2018 & 11.26 & 1.31 & 100 \\ 
political-bots-2019 & 0 & 20.60 & 100 \\ 
varol-2017 & 83.17 & 72.28 & 100 \\ 
verified-2019 & 88.60 & 98.15 & 100 \\ 
reddit & 0 & 0 & 100 \\ \hline
Average & 55.04$\pm$\newline 31.32 & 51.61$\pm$\newline34.49 & 100\\
\hline
\end{tabular}
\caption{Percentage of dataset analyzed by each algorithm.}
\label{tab:percanalyzed}
\end{table}

We present the results of BotBuster-Singular and BotBuster-Full in Table \ref{tab:modelresults}. 
In general, singular runs of each dataset (BotBuster-Singular) perform better than the baselines, but does not perform as well as the merged training model (BotBuster-Full). Thus, augmenting the training data by merging all the datasets for model training and hyperparamter tuning enhances the model performance.

Out of the three model variations trained, the BotBuster-4 variation performed the best (F1=72.23$\pm$18.84) across all the test datasets. It outperforms the baselines of BotHunter (F1=49.61$\pm$30.10) and Botometer (F1=40.63$\pm$26.33). From the independent dataset analysis, we observe that the standard deviation of F1 scores of BotBuster is smaller than the baseline algorithms. BotBuster classifier is thus less susceptible to cross dataset variances.

Observing that BotBuster-3/4 variations performed better than BotBuster-1/2 variations lends to the fact that a post-level post expert differentiates post information between bots and humans better than an account-level post expert. 
While the use of derived post linguistic values does not significantly improve model accuracy for account-level post expert (BotBuster-1 vs BotBuster-2), it does so for the post-level post expert (BotBuster-3 vs BotBuster-4).
BotBuster performs most poorly are datasets from 2017, suggesting the mutable behavior of bots over time \cite{Luceri_Deb_Giordano_Ferrara_2019}.

Our input-agnostic BotBuster architecture requires only specification of matching field and does not rely on any specific data format and opens up possibilities for multi-platform bot-detectors. As such, BotBuster is able to perform predictions on Reddit data, despite it not having the same fields as Twitter data. 
Observing that BotBuster-4 performs well on both the Reddit and Twitter datasets supports our architecture and opens up discussion for the similar behaviour of bots on both platforms.

\begin{table*}[!htbp]
\centering
\begin{tabular}{|l|p{2cm}p{2cm}p{2cm}p{2cm}p{2cm}p{2cm}|}
\hline
\textbf{Dataset} & \textbf{BotHunter} & \textbf{Botometer} & \textbf{BotBuster-Singular} & \textbf{BotBuster-Full} & \textbf{BotBuster-Subset} \newline (BotHunter returns) & \textbf{BotBuster-Subset} \newline (Botometer returns) \\ \hline
\multicolumn{7}{|l|}{\textbf{BotBuster-1: username, screenname, descriptions, user metadata experts + account-level post expert}} \\ 
astroturf & 15.60 & 33.50 & \textbf{50.00} &\textbf{65.90} & \underline{15.72} & \underline{33.60} \\ 
botometer-feedback-2019 & 74.10 & 53.68 & 49.52 & 29.10 & 25.19 & 25.20 \\
botwiki-2019 & 53.13 & 92.89 & 45.31 & \underline{79.03} & \underline{79.00} & 79.02 \\
cresci-rtbust-2019 & 62.90 & 60.10 & \underline{61.84} & 51.78 & 51.78 & 51.78 \\
cresci-stock-2018 & 37.36 & 38.12 & \textbf{61.22}& \textbf{71.49} & \underline{56.17} & \underline{54.53}\\
gilani-2017 & 63.91 & - & 53.62 & 47.88 & 52.55 & - \\
midterm-2018 & 15.30 & 11.90 & \textbf{46.64} & \textbf{84.80} & 8.00 & 16.06 \\
political-bots-2019 & - & 20.60 & \textbf{45.45} & \textbf{98.38} & - & \underline{95.03}\\
varol-2017 & 73.80 & 65.30 & 44.74 & 32.18 & 32.89 & 33.40\\
verified-2019 & 100 & 30.20 & \underline{46.80} & \underline{87.10} & 97.29 & \underline{80.34}\\
reddit & - & - & 43.51 & 53.21 & - & - \\ 
Average & 55.12 & 45.14 & \underline{49.87}& \textbf{63.71} & 46.51 & \underline{52.11} \\
\hline
\multicolumn{7}{|l|}{\textbf{BotBuster-2: username, screenname, descriptions, user metadata experts + account-level post expert with derived post values}} \\ 
astroturf & 15.60 & 33.50 & \textbf{43.13} & \textbf{68.24} & \underline{16.50} & \underline{33.54} \\ 
botometer-feedback-2019 & 74.10 & 53.68 & 22.05 & 29.11 & 29.23 & 25.20\\
botwiki-2019 & 53.13 & 92.89 & 44.88 & \underline{79.12} & \underline{78.93} & 79.36\\
cresci-rtbust-2019 & 62.90 & 60.10 & 51.32 & 51.78 & 51.67 & 51.00\\
cresci-stock-2018 & 37.36 & 38.12 & \textbf{65.89} & \textbf{71.50} & 56.17 & \underline{54.53} \\
gilani-2017 & 63.91 & - & 50.04 & 47.88 & 52.55& -\\
midterm-2018 & 15.30 & 11.90 & \textbf{66.14} & \textbf{84.80}& 8.02 & \underline{16.00}\\
political-bots-2019 & - & 20.60 & \textbf{97.45} & \textbf{98.38} & - & \underline{99.23}\\
varol-2017 & 73.80 & 65.30 & 42.02 & 32.20 & 32.90 & 33.40\\
verified-2019 & 100 & 30.20 & \underline{47.36} & \textbf{87.05} & 98.60 & \underline{88.40}\\
reddit & - & - & 48.32 & 57.02 & - & - \\ 
Average & 55.12 & 45.14 & \underline{52.60} & \textbf{64.28} & 47.17 & \underline{53.41}\\
\hline
\multicolumn{7}{|l|}{\textbf{BotBuster-3: username, screenname, descriptions, user metadata experts + post-level post expert} }\\ 
astroturf & 15.60 & 33.50 & \textbf{50.34} & \textbf{80.96} &  \underline{18.90} & \underline{29.95}  \\ 
botometer-feedback-2019 & 74.10 & 53.68 & 28.30 & 29.67 & 25.46 & 29.23\\
botwiki-2019 & 53.13 & 92.89 & 38.57 & \underline{80.96} & \underline{78.93} & 79.51\\
cresci-rtbust-2019 & 62.90 & 60.10 & 42.32 & 51.57 & 52.10 & 51.66\\
cresci-stock-2018 & 37.36 & 38.12 & \textbf{70.33} & \textbf{71.49} & \underline{56.17} & \underline{54.53}\\
gilani-2017 & 63.91 & - & 55.09 & 48.00 & - & 52.56 \\
midterm-2018 & 15.30 & 11.90 & \textbf{83.89} & \textbf{84.87} & 8.10 & 6.26 \\
political-bots-2019 & - & 20.60 & \textbf{52.36} & \textbf{49.59} & - & \underline{97.95}\\
varol-2017 & 73.80 & 65.30 & 47.80 & 32.18 & 32.89 & 33.40\\
verified-2019 & 100 & 30.20 & \underline{47.37} & \textbf{87.00} & \underline{97.29} & \underline{88.44}\\
reddit & - & -  & 52.87 & 57.03 & - & -\\ 
Average & 55.12 & 45.14 & \underline{51.74} & \textbf{59.39} & 46.23 & \underline{52.35} \\
\hline
\multicolumn{7}{|l|}{\textbf{BotBuster-4: username, screenname, descriptions, user metadata experts + post-level post expert with derived post values} }\\ 
astroturf & 15.60 & 33.50 & \textbf{58.00} & \textbf{76.90} & \underline{17.21*} & \underline{34.28*} \\ 
botometer-feedback-2019 & 74.10 & 53.68 & 41.82 & \underline{54.20} & 68.62* & \underline{59.90*}\\
botwiki-2019 & 53.13 & 92.89 & \underline{70.00} & \underline{80.90} & \underline{78.90*} & \underline{79.51*}\\
cresci-rtbust-2019 & 62.90 & 60.10 & \textbf{63.14} & \textbf{67.20} & \underline{72.23*} & \underline{70.20*} \\
cresci-stock-2018 & 37.36 & 38.12 & \textbf{62.19} & \textbf{79.35} & \underline{75.44}& \underline{71.13}\\
gilani-2017 & 63.91 & - & 57.04 & 48.80 & \underline{53.23*} & - \\
midterm-2018 & 15.30 & 11.90 & \textbf{83.93} & \textbf{94.13} & \underline{90.20*} & \underline{73.48*}\\
political-bots-2019 & - & 20.60 & \textbf{100} & \textbf{98.38} & - & \underline{97.95*}\\
varol-2017 & 73.80 & 65.30 & 41.83 & 45.50& 48.78* & 51.39*\\
verified-2019 & 100 & 30.20 & \textbf{100} & \underline{89.15} & \underline{99.72} & \underline{90.62}\\
reddit & - & - & 69.77 & 60.04 & - & - \\ 
Average & - & - & 67.97 & 72.23 & - & 69.83\\ 
Avg (BotHunter returns) & 55.12 & - & 64.21 & 63.61 & 67.14 & - \\ 
Avg (Botometer returns) & - & 45.14 & 70.68 & 76.19 & - & 69.83\\ 
\hline
\end{tabular}
\caption{Macro-F1 scores for three variations of BotBuster and the baseline comparisons. BotBuster results better than all baselines are \textbf{bolded} and results better than only one baseline are \underline{underlined}. For BotBuster-4, we also report the average BotBuster scores compared to the datasets the baselines return results in addition to the overall average score, and an * means a significant difference in results at the $p<0.05$ level.}
\label{tab:modelresults}
\end{table*}

\paragraph{Known Information Expert}
On average, our datasets have 16\% of accounts containing known information, indicating the merit of incorporating knowledge about the social media platforms. 
The breakdown of accounts detected through known information is shown in Table \ref{tab:datasets}. These information should be updated as platforms improve their service and include more indicators.

\paragraph{Evaluation against external dataset}
The results of the external evaluation against the TwiBot-20 dataset are presented in Table \ref{tab:exteval}. BotBuster outperforms the baseline models (F1=60.92 vs avg $F1$=34.94), illustrating the robustness of the model in characterizing bots.

\begin{table}[!htbp]
\centering
\begin{tabular}{|p{3cm}|p{3cm}p{1.5cm}|}
\hline
\textbf{Model} & \textbf{Users analyzed (\%)} & \textbf{Macro-F1} \\ \hline
BotHunter & 99.15 & 49.02\\
Botometer & 91.38 & 20.86 \\ \hline
BotBuster-4 & \textbf{100} & \textbf{60.92}\\
BotHunter returns & 99.15 & 56.97\\
Botometer returns & 91.38 & 52.89\\ \hline
TwiBot-20 & 100 & 85.46 \\
\hline
\end{tabular}
\caption{Macro-F1 scores of external evaluation dataset}
\label{tab:exteval}
\end{table}

\begin{figure}
\centering
\includegraphics[width=0.5\textwidth]{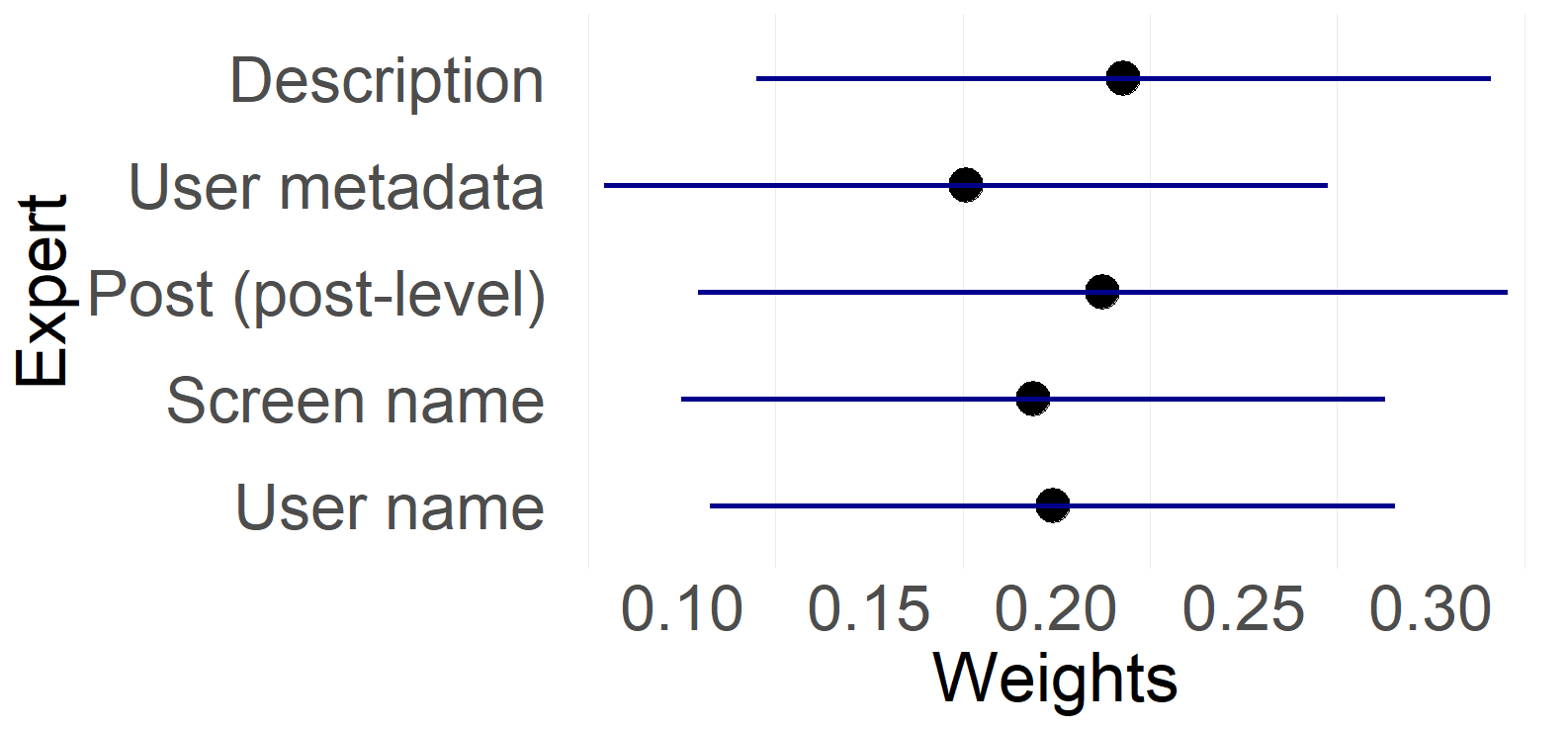} 
\caption{Distribution of expert importance of BotBuster-4 architecture}
\label{fig:expertweights}
\end{figure}

\paragraph{Understanding expert importance}
$\bm{P}(\text{bot})$ is a weighted sum of the experts. Figure \ref{fig:expertweights} summarize the expert weights for all combinations of data input to BotBuster-4 derived by the Expert Importance Gating network. A higher value reflects a higher importance for the expert. 

The input weights are almost evenly distributed across the experts, suggesting an almost equal importance on each expert for final prediction. However, there is more emphasis on post information and descriptions and lesser value on screennames and usernames, which suggests that longer writing is a key determinant of $\bm{P}(\text{bot})$. Screennames and usernames are typically single words. In our dataset, the average user name length is $3.02\pm5.15$ characters, and the average screen name length is $3.28\pm6.23$ characters. The average Levenshtein distance between both values across our dataset is $2.36\pm4.89$, suggesting users typically use similar strings for both these names. An account's description is more lengthy, usually a sentence with $3.62\pm6.33$ words, and the maximum allowed tweet length ranges from 40-70 words. The gating network places more emphasis on text input pillars compared to metadata, and particularly on longer text-input pillars that contain more information.

\section{Discussion}
In this work, we built a social bot detection model, BotBuster, leveraging on the mixture of experts architecture. 

\paragraph{Generalizable feature engineering and model design}
We performed the same feature engineering on 10 Twitter datasets and one Reddit dataset, demonstrating that the features are generalizable across datasets and platforms. 
These features are general to the social media space, because every user in all social media platforms have a username, posts and other account metadata. 
In the event any of the feature is missing because the platform does not have the feature tagged to the user, the expert is not used in the evaluation to whether the user is a bot or not.
Thus, this method is scalable to general bot detection and only requires a mapping of feature names.

Our results identify that a merged training strategy performs better than individual dataset training strategies. 
Training the model on an augmented dataset aids in model generalizability where the model learns features across a variety of data types.
BotBuster demonstrated its robustness in achieving a 2-4 times better accuracy than the baseline algorithms on an external dataset. 
Although BotBuster was trained on older datasets, it is sufficiently robust enough for bot detection on the newer TwiBot-20 dataset. 
The accuracy achieved by BotBuster is lower than the model built by the TwiBot-20 authors, which may be due to the use of network information in prediction.
Given that BotBuster can detect newer bot styles reasonably well, the generalizability and performance of BotBuster seems acceptable.

\paragraph{Difficulty of differentiating bot and humans}
We concatenate all features used as input to BotBuster into vectors and perform Principal Component Analysis before visualizing the top 2 components. Through this, we discern the level of feature separation between bots and humans. In datasets where BotBuster performs poorly, we observe poor separation. This is consistent with past work
where the same datasets are do not achieve high performance in generalized bot detectors \cite{yang2020scalable}.

\begin{figure*}
\centering
\includegraphics[width=1.0\textwidth]{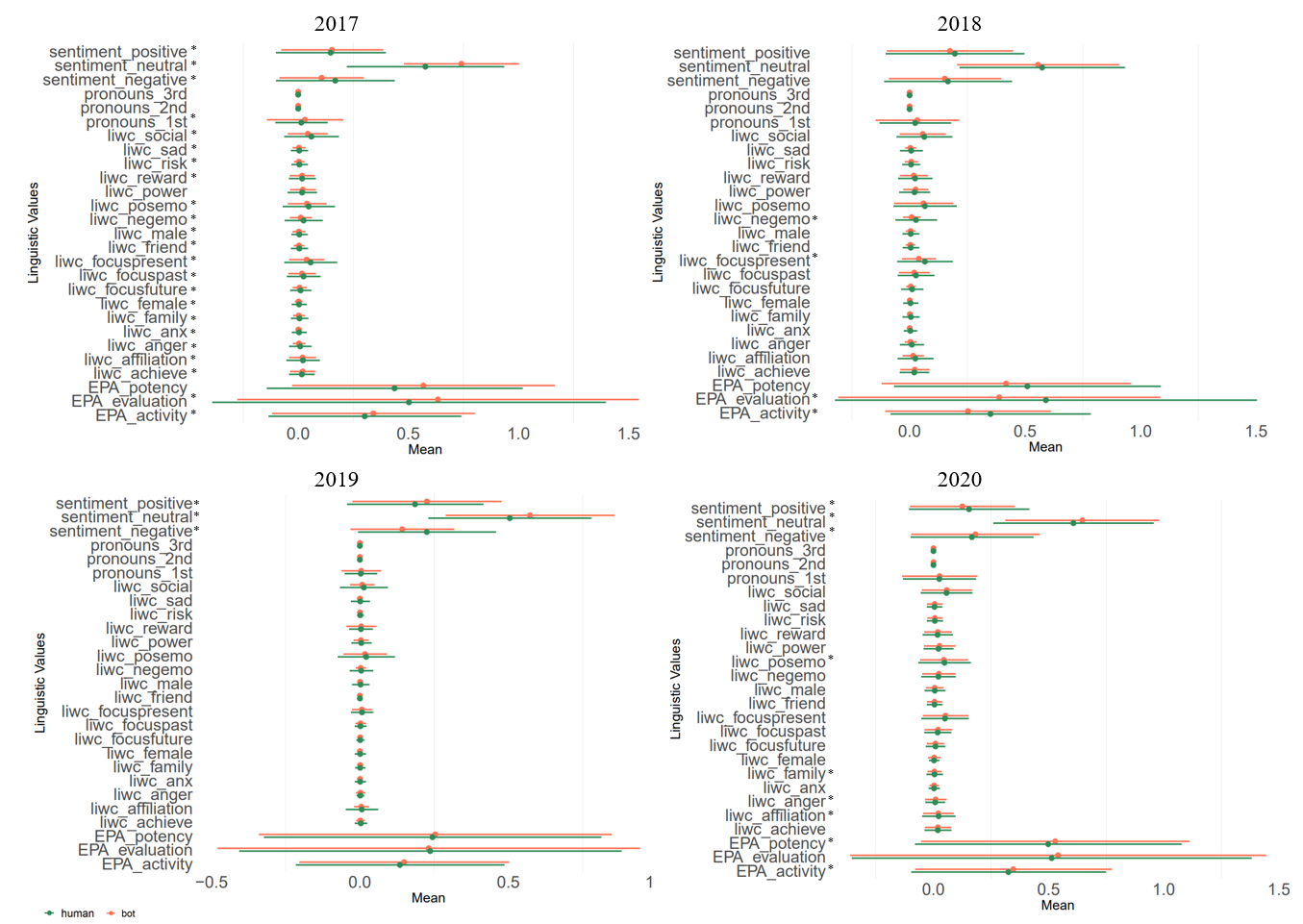} 
\caption{Change in linguistic features over years of bot datasets. * denotes significant difference ($p<0.05$) between bot/human}
\label{fig:linguisticchange}
\end{figure*}

\paragraph{The changing nature of bots}
BotBuster performs best on datasets curated from 2018-2020 and worst on datasets curated prior to 2017. 
It also does not perform as well as the newer bot detection model, TwiBot-20, which was developed on the 2020 dataset. 
This alludes to the changing nature of bot account features. Figure \ref{fig:linguisticchange} plots the the difference in linguistic values in bot/human accounts in our datasets across the years, signalling changes in writing styles. In 2017, bot and human accounts have many significantly different features as compared to the following years. The reduction and change in significantly different features increases the challenge of developing lasting bot detection algorithms.
We note that this analysis is restricted to the datasets in our study, although the datasets are widely used.


\paragraph{Stability of BotBuster}
One key characteristic of a good bot detection algorithm is the stability of its Bot Probability Score. A stable bot score changes minimally across an investigation time frame, thus providing reliable characterization of bots for downstream tasks like detection of influence campaigns \cite{bothunterstability}. 
\begin{figure}[t]
\centering
\includegraphics[width=0.6\columnwidth]{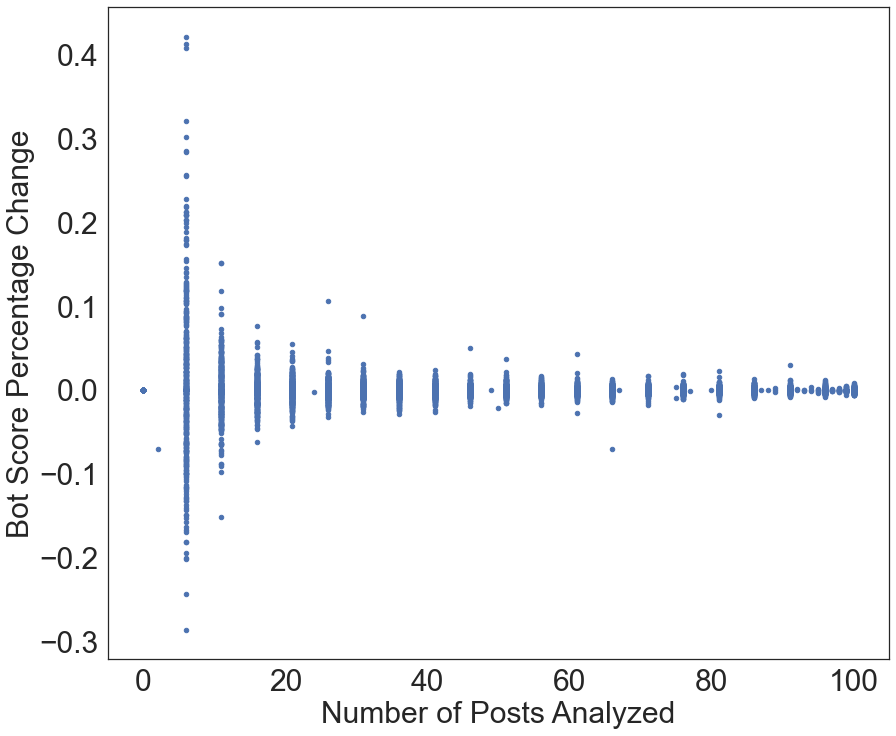} 
\caption{Change in BotBuster-4 scores}
\label{fig:botscorestability}
\end{figure}

We empirically study the amount of data required for a stable BotBuster score.
We randomly select 500 still-alive bots and 500 non-bots from the dataset and collected their latest 100 posts using the Twitter V2 REST API. We then run the BotBuster-4 algorithm beginning with one post then by incremental steps of 5 posts, up to 100 posts. 
We analyze the percentage change in BotBuster score across the number of posts (Figure \ref{fig:botscorestability}). 
The difference in scores initially changes drastically from an initial change of $-0.286\pm0.0871$, then drops to a change of $3.70\text{E-}5\pm1.15\text{E-}2$ at 21 posts, and tends to 0 after 36 posts ($-7.80\text{E-6}\pm6.86\text{E-3}$). 
Thus, BotBuster scores do stabilize, lending confidence in the algorithm.
The observations provide evidence for usage of BotBuster estimation: at least 36 posts should be collected for a score that changes minimally.

\begin{figure*}
\centering
\includegraphics[width=1.0\textwidth]{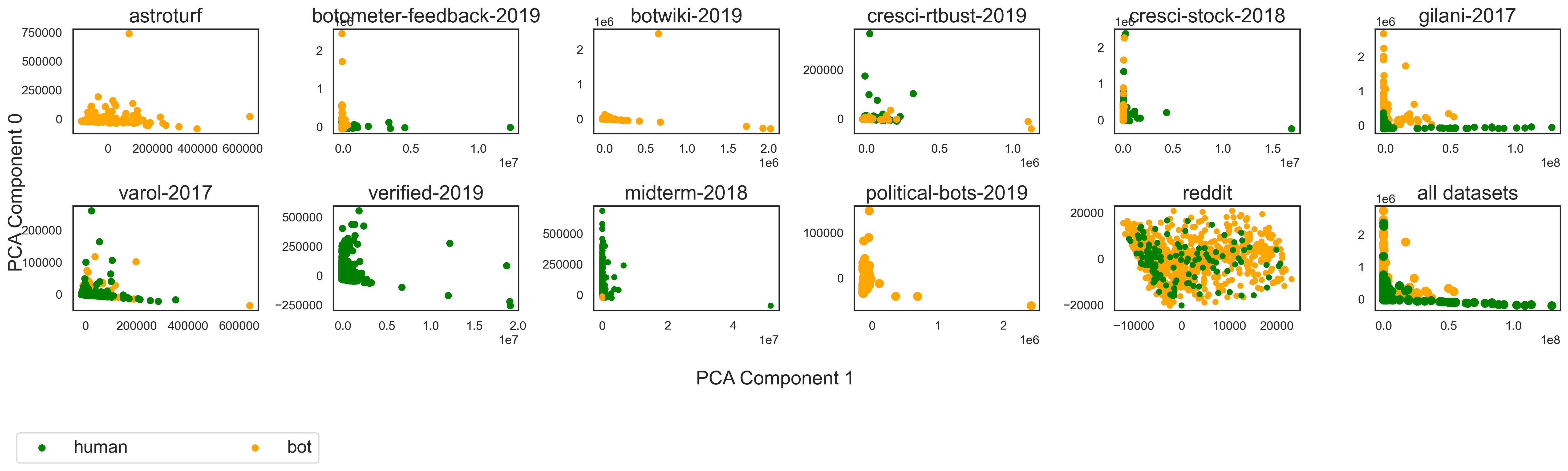} 
\caption{PCA Plot of datasets show the difficulty of bot detection and provide clues to the performance of BotBuster}
\label{fig:pcaplot}
\end{figure*}

\paragraph{Limitations and Future Work}
The changing nature of bot characteristics requires continual research to continually update and build new algorithms. The supervised learning construct of BotBuster relies on human annotations of accounts as a supervised learning approach.
Additionally, the bot/human labels are manually annotated, meaning there could be false positives in either of the classes where humans are unable to distinguish bot and human account.
Future work can exploit network structure or temporal activity as additional experts. It can explore incorporating human expertise through a feedback mechanism.


\paragraph{Ethical Considerations}
Social bot detection through automated means bring about a key ethical consideration: accuracy, transparency and robustness of a social bot detection algorithm collectively forms a ``devil's triangle" \cite{thieltges2016devil}. Accuracy is paramount as misclassification can lead to the deplatforming of legitimate social media users. At the same time, a positive bot classification does not indicate a malicious account users should further discern the account characteristics to be sure to weed out malicious accounts only.
To enable wider usage, the algorithm should be transparent. However, an increase in algorithm transparency provides bot-operators information to adapt bot account characteristics to evade detection, increasing the variation of bot characteristics. The alteration of bot behavior based on the knowledge of the bot detection algorithm creates a drop in the robustness and accuracy of the algorithm in the detection of new and evolved bot accounts. Bot detection is a cat-and-mouse game; transparency must be balanced with robustness  \cite{fazzolari2020emotions}. All three pillars must be balanced because excessive focus on any of them can give the reign of social media space to malicious bot operators.

\section{Conclusion}
Given the prevalence of bots on social media platforms and the possibility of incomplete data collection, improved platform API data formats, there is a the need for an improved bot detection method to be format-agnostic and handle incomplete data. 
Bot detection is still an open research problem, as our analysis show bot/human differentiation can be difficult and bot features change over time.
We develop BotBuster, a novel mixture of experts bot detection approach that handles incomplete data collection. 
Its performance on cross-platform datasets gives hope to generalizability of bot detectors and suggests that bots operate similarly across these two platforms.

\section*{Acknowledgments}
Removed for anonymous submission

{\small
\bibliography{aaai22}}

\appendix
\begin{table*}[!htbp]
\centering
\begin{tabular}{|l|l|l|}
\hline
\textbf{Feature} & \textbf{Type} & \textbf{Reference library} \\ \hline
\multicolumn{3}{|l|}{\textbf{Extracted User Features}} \\ 
username & string & - \\
screenname & string & - \\ 
description & string & - \\ 
number of followers & integer & - \\ 
number of following & integer & - \\ 
total number of posts & integer & - \\ 
listed count & integer & - \\ 
is verified & boolean & - \\ 
is protected & boolean & - \\ 
\hline
\multicolumn{3}{|l|}{\textbf{Extracted Post Features}} \\ 
post text & string & - \\ 
number of retweets & integer & - \\
number of likes & integer & - \\ 
number of quotes & integer & - \\ 
number of replies & integer & - \\ 
\hline
\multicolumn{3}{|l|}{\textbf{Derived Post Features}} \\ 
sentiment & float & \cite{perez2021pysentimiento} \\ 
reading score & float & \cite{10.2307/43086724} \\ 
EPA scores & integer & \cite{smith2016mean,tyagi2021challenges} \\
LIWC features: time, affect, social, drives, pronouns values & integer & \cite{doi:10.1177/0261927X09351676} \\ 
\hline
\end{tabular}
\caption{List of features used in BotBuster}
\label{tab:features}
\end{table*}

\begin{table*}[!htbp]
\centering
\begin{tabular}{|l|p{0.8cm}|p{1.2cm}|p{2.7cm}||p{0.8cm}|p{0.8cm}|p{0.8cm}|p{0.8cm}|p{1cm}|p{1.6cm}|}
\hline
\textbf{Dataset} & \textbf{Bots} & \textbf{Humans} & \textbf{Reference} & \textbf{User name} & \textbf{Screen name} & \textbf{Desc} & \textbf{Posts} & \textbf{User Meta data} & \textbf{Known Information (\%)}\\ \hline

astroturf & 585 & 0 & \cite{sayyadiharikandeh2020detection} & \cellcolor{Gray} & \cellcolor{Gray} & \cellcolor{Gray} & \cellcolor{Gray} & \cellcolor{Gray} & 0.3\\ 

botometer-feedback-2019 & 143 & 386 & \cite{yang2019arming} &  &  & \cellcolor{Gray} & \cellcolor{Gray} & \cellcolor{Gray} & 7.0\\ 

botwiki-2019 & 704 & 0 & \cite{yang2020scalable} & \cellcolor{Gray} & \cellcolor{Gray} & \cellcolor{Gray} & \cellcolor{Gray} & \cellcolor{Gray} & 49\\ 

cresci-rtbust-2019 & 391 & 368 & \cite{mazza2019rtbust} & \cellcolor{Gray} & \cellcolor{Gray} & \cellcolor{Gray} & \cellcolor{Gray} & \cellcolor{Gray} & 0.53\\ 

cresci-stock-2018 & 18508 & 7479 & \cite{cresci2018fake} & \cellcolor{Gray} & \cellcolor{Gray} & \cellcolor{Gray} & \cellcolor{Gray} & \cellcolor{Gray} & 0.058\\ 

gilani-2017 & 1130 & 1522 & \cite{diesner2017proceedings} & \cellcolor{Gray} & \cellcolor{Gray} & \cellcolor{Gray} & \cellcolor{Gray} & \cellcolor{Gray} & 28 \\ 

midterm-2018 & 42446 & 8092 & \cite{yang2020scalable} &  &  & \cellcolor{Gray} & \cellcolor{Gray} & \cellcolor{Gray} & 0.93\\ 

political-bots-2019 & 62 & 0 & \cite{yang2019arming} & & & \cellcolor{Gray} & \cellcolor{Gray} & \cellcolor{Gray} & 1.6\\ 

varol-2017 & 826 & 1747 & \cite{varol2017online} & \cellcolor{Gray} & \cellcolor{Gray} & \cellcolor{Gray} & \cellcolor{Gray} & \cellcolor{Gray} & 1.7\\ 

verified-2019 & 0 & 2000 & \cite{yang2020scalable} & \cellcolor{Gray} & \cellcolor{Gray} & \cellcolor{Gray} & \cellcolor{Gray} & \cellcolor{Gray}& 1.3\\

reddit & 500 & 167 & - &  & \cellcolor{Gray} & \cellcolor{Gray} &  & \cellcolor{Gray} & 87\\ \hline

Total & 65295 & 21761 & - & \cellcolor{Gray} & \cellcolor{Gray} & \cellcolor{Gray} & \cellcolor{Gray} & \cellcolor{Gray} & 16$\pm$28\\ 

\hline
\hline
TwiBot-20 \newline (external evaluation) & 640 & 543 & \cite{feng2021twibot} &  &  &  & \cellcolor{Gray} & & 28\\ 
\hline
\end{tabular}
\caption{Dataset composition. The datasets are hydrated in July 2021 and the greyed-out cells indicate where fields with incomplete data. We also list the percentage of accounts where there is known information and thus processed by the Known Information expert.}
\label{tab:datasets}
\end{table*}



\end{document}